\documentclass[pra,aps,10pt,twocolumn,showpacs,amsmath,amssymb]{revtex4-1}

\usepackage[english]{babel}
\usepackage{afterpage}
\usepackage{graphicx}
\usepackage{dcolumn}
\usepackage{bm}
\usepackage{color}
\usepackage{slashed}
\usepackage{latexsym}

\DeclareMathAlphabet\mathbfcal{OMS}{cmsy}{b}{n}

\newcommand{\dd}{d}  
\newcommand{\ii}{i}  
\newcommand{\ee}{e}  

\newcommand{\op}[1]{\hat{#1}} 
\newcommand{\bra}[1]{\langle{#1}\vert} 
\newcommand{\ket}[1]{\vert{#1}\rangle} 

\begin{document}

\title{Diffraction at a time grating in electron-positron pair creation from vacuum}
\author{J. Z. Kami\'nski$^1$}
\author{M. Twardy$^2$}
\author{K. Krajewska$^{1}$}
\email[E-mail address:\;]{Katarzyna.Krajewska@fuw.edu.pl}

\affiliation{$^1$Institute of Theoretical Physics, Faculty of Physics, University of Warsaw, Pasteura 5,
02-093 Warszawa, Poland\\
$^2$Faculty of Electrical Engineering, Warsaw University of Technology, Plac Politechniki 1, 00-661 Warszawa, Poland}
\date{\today}

\begin{abstract}
The Sauter-Schwinger process of electron-positron pair creation from vacuum, driven by a sequence of time-dependent electric-field pulses, is studied in the framework 
of quantum-field theoretical approach. As demonstrated by our numerical results, the probability distributions of produced pairs exhibit intra- and inter-pulse 
interference structures. We show that such structures can be observed beyond the regime of applicability of the WKB theory, which was the focus of earlier investigations. 
Going beyond these developments, we perform the analysis of the time-evolution operator for an arbitrary eigenmode of the fermionic field. This shows that a perfect
coherent enhancement of the inter-pulse peaks can never be reached. A nearly perfect coherence, on the other hand, is due to nonadiabatic transitions at avoided 
crossings of the phases defining the unitary time evolution. This analysis allows us to determine the conditions under which the nearly perfect coherence is lost.

\end{abstract}

\maketitle

\section{Introduction}
\label{sec::intro}

The quantum vacuum is one of the most exciting subjects of contemporary physics. While the vacuum instability in the presence of a static electric field, which results in
electron-positron ($e^-e^+$) pair creation, has been predicted decades ago~\cite{Sauter,Heisenberg-Euler,Schwinger}, there is no its direct experimental confirmation yet. 
The reason being that the effect is very weak and, even then, an enormous (for laboratory conditions) electric field is necessary to expel real pairs out of the vacuum.

A number of proposals have been put forward aiming at enhancing the signal of Sauter-Schwinger pairs. This, essentially, is done by tailoring 
the external electric field. In this respect, the dynamically assisted mechanism was proposed~\cite{Schutzhold,Dunne1,Orthaber,Fey,Jansen,Akal,Otto1,Otto2}, in which the pair signal is
enhanced by orders of magnitude when superposing a slowly varying in time but strong with a rapidly oscillating but weak electric fields. Other field configurations were also
considered including a combination of a static and alternating electric fields~\cite{Jiang} or a combination of three electric fields oscillating in time at different 
scales~\cite{Sitiwaldi}. All these studies show that the $e^-e^+$ pair creation is sensitive to details of the external field
configuration. Hence, raising the question of optimal control of the process~\cite{Hebenstreit,Fillion}.

The most relevant to our work is the idea pursued by Akkermans and Dunne~\cite{Akkermans}, later on followed also by Li and collaborators~\cite{Li1,Li2} for boson pair creation,
where a sequence of identical time-dependent electric field modulations was considered. This resulted in a multislit interference pattern in the momentum distribution 
of created particles. It was demonstrated in~\cite{Akkermans} that such interference occurs for a sequence of $N$ alternating-sign modulations, with the central 
value scaling as $N^2$ compared to the maximum distribution originating from a single modulation. This was supported by a comparison with an analytically predicted
$N^2$-enhancement law arising from the WKB theory and the turning point analysis~\cite{Dumlu1,Dumlu2}. 
Our purpose now is to develop the aforementioned idea of Akkermans and Dunne~\cite{Akkermans} such that it goes beyond the regime of applicability of the WKB theory and it relates exclusively
to a train of $N_{\rm rep}=N/2$ identical electric field pulses. Thus, it is also different than a modulated pulse train analyzed in Ref.~\cite{Xie}.

Note that for an electric field pulse, the conditions~\eqref{a1} and~\eqref{a2} below are satisfied~\cite{Becker}, which is not the case for a single field
modulation considered in~\cite{Akkermans}. More detailed theoretical formulation of our problem is given in Sec.~\ref{theory} (see, also Appendix~\ref{Vlasov}),
which to large extent follows the derivation from~\cite{Grib1,Grib2}. Nevertheless, we present it for convenience of the reader. Next, in Sec.~\ref{distributions},
the momentum distributions of created particles are demonstrated, exhibiting intra- and inter-pulse interference patterns. The latter show a nearly perfect 
$N_{\rm rep}^2$-enhancement with respect to the former one, meaning that all major inter-pulse peaks (not just the central one)  
scale approximately like $N_{\rm rep}^2$ with respect to the intra-pulse modulations. As we demonstrate, this happens for the electric field
parameters for which the WKB approximation is not applicable, calling for a different interpretation of the observed patterns than the one offered in~\cite{Akkermans}.
This is addressed in Sec.~\ref{evolution} by studying the unitary time-evolution matrix of an arbitrary eigenmode of the fermionic field (see, also Appendix~\ref{appendix1}). 
More precisely, we analyze the functional 
dependence of the phases defining the time evolution ($\vartheta_1$ and $\vartheta_2$) on the asymptotic particles momenta. We observe that at the given values of momenta, 
for which adiabatic transitions between both phases occur due to their avoided crossings, very pronounced peaks appear. As we argue, these peaks are nearly perfectly coherent 
but can be diminished by increasing the gap at the avoided crossings. There are also actual crossings of $\vartheta_1$ and $\vartheta_2$, at which the signal of pair creation is zero.
Note that our interpretation is independent of the regime of parameters and it explains in detail the properties of the probability distributions of created particles;
thus, it complements the previous investigations~\cite{Akkermans,Li1,Li2}. Our closing remarks are given in Sec.~\ref{conclusions}.

Throughout the paper, we keep $\hbar=1$. However, in our numerical analysis we use relativistic units (rel. units) 
such that $\hbar=m_{\rm e}=|e|=c=1$, where $m_{\rm e}$ is the electron rest mass and $e<0$ is its charge. Also, we employ the Feynman notation $\slashed{a} =\gamma^{\mu} a_{\mu}$ for the contraction 
with the Dirac matrices $\gamma^{\mu}$. For the relativistic scalar product, we use the signature $(+---)$.

\section{Theoretical formulation}
\label{theory}

We consider the electron-positron pair creation from vacuum by a homogeneous in-space, time-dependent electric field which is described by the four-vector 
potential $A^\mu(x)=(0,{\bm A}(t))\equiv(0,0,0,A(t))$, with an arbitrary $A(t)$ such that 
\begin{equation}
\underset{t\rightarrow-\infty}{\lim}A(t)=\underset{t\rightarrow+\infty}{\lim}A(t).
\label{a1}
\end{equation}
Thus, a pulsed electric field which oscillates linearly along the $z$-direction, $c{\cal F}^{\mu 0}(x)=(0,{\mathbfcal{E}}(t))=(0,0,0,{\cal E}(t))$, with
\begin{equation}
\int_{-\infty}^{+\infty}\dd t\,{\cal E}(t)=0,
\label{a2}
\end{equation}
is considered. (${\cal F}^{\mu\nu}(x)$ above is the electromagnetic field tensor.) The last condition arises as ${\cal E}(t)=-\frac{\dd A(t)}{\dd t}$. Our purpose now is to construct one particle
solutions of the Dirac equation in such field. 

\subsection{One particle solutions of the Dirac equation}
\label{one-particle}

The Dirac equation coupled to the pulsed electric field has the form
\begin{equation}
\bigl(\ii\slashed{\partial}-e\slashed{A}-m_{\rm e}c\bigr)\Psi(x)=0.
\label{t1}
\end{equation} 
Following Ref.~\cite{Schweber}, we transform this equation into a second order differential equation by assuming that there exists a bispinor $\chi(x)$ such that
\begin{equation}
\Psi(x)=\bigl(\ii\slashed{\partial}-e\slashed{A}+m_{\rm e}c\bigr)\chi(x).
\label{t2}
\end{equation}
Combining the two equations we find out that $\chi(x)$ solves
\begin{equation}
\bigl[\slashed{\partial}^2+\ii e (\slashed{\partial}\slashed{A}+\slashed{A}\slashed{\partial})-e^2\slashed{A}^2+(m_{\rm e}c)^2\bigr]\chi(x)=0.
\label{t3}
\end{equation}
Note that this equation offers twice that many solutions as the Dirac equation itself. Thus, in order to have one-to-one correspondence between both sets of solutions,
we need to narrow down the number of solutions of Eq.~\eqref{t3}. The way it is done here reduces the problem to solving a differential equation
for a single scalar function.

To demonstrate this, we note first that the problem is translationally invariant. Hence, it is justified to look for the bispinor $\chi(x)$ in the form,
\begin{equation}
\chi(x)=\ee^{\ii {\bm p}\cdot {\bm x}}\chi_{{\bm p}}(t),
\label{t4}
\end{equation}
where $\chi_{{\bm p}}(t)$ is independent of the position ${\bm x}$ and we label it by an asymptotic momentum of a particle ${\bm p}$. With this substitution
and accounting for the fact that the electric field oscillates in the $z$-direction, Eq.~\eqref{t3} becomes
\begin{equation}
\Bigl[\frac{\dd^2}{\dd t^2}+\ii ce{\cal E}(t)\gamma^0\gamma^3+\omega_{\bm p}^2(t)\Bigr]\chi_{\bm p}(t)=0.
\label{t5}
\end{equation}
Here,
\begin{equation}
\omega_{\bm p}^2(t)=c^2{\bm p}_\perp^2+c^2(p_{\|}-eA(t))^2+(m_{\rm e}c^2)^2
\label{t6}
\end{equation}
is expressed in terms of the longitudinal $p_{\|}$ and the transverse ${\bm p}_\perp$ components of the particle asymptotic momentum, which are defined as
\begin{equation}
p_{\|}={\bm p}\cdot{\bm e}_z,\quad {\bm p}_\perp={\bm p}-p_{\|}{\bm e}_z.
\label{t7}
\end{equation}
Eq.~\eqref{t5} is further simplified assuming that $\chi_{\bm p}(t)$ remains an eigenstate of the matrix $\gamma^0\gamma^3$. 
Actually, $\gamma^0\gamma^3$ has two doubly degenerate eigenvalues $\pm 1$. It turns out, however, that it is enough to choose one of them~\cite{Grib1,Grib2}.
Specifically, we shall keep in the following
\begin{equation}
\chi_{\bm p}(t)\equiv\chi_{{\bm p}\lambda}(t)=\psi_{\bm p}(t)u_\lambda,
\label{t8}
\end{equation}
where $\gamma^0\gamma^3 u_\lambda=u_\lambda$. Hence, for as long as
\begin{equation}
u_+=\frac{1}{\sqrt{2}}\begin{pmatrix}
1 \\ 0 \\ 1 \\ 0
\end{pmatrix},\quad
u_-=\frac{1}{\sqrt{2}}\begin{pmatrix}
0 \\ -1 \\ 0 \\ 1
\end{pmatrix},
\label{t9}
\end{equation}
(meaning that $\lambda=\pm$) the problem simplifies to solving a differential equation for a scalar function $\psi_{\bm p}(t)$,
\begin{equation}
\Bigl[\frac{\dd^2}{\dd t^2}+\ii ce{\cal E}(t)+\omega_{\bm p}^2(t)\Bigr]\psi_{\bm p}(t)=0.
\label{a9}
\end{equation}
Finally, we also note that $u_\lambda^\dagger u_{\lambda'}=\delta_{\lambda\lambda'}$.

Let us now interpret the resulting solutions.
It follows from~\eqref{a9} that, in the remote past ($t\rightarrow-\infty$), the scalar function $\psi_{\bm p}(t)$ satisfies the asymptotic equation,
\begin{equation}
\Bigl[\frac{\dd^2}{\dd t^2}+\omega_{\bm p}^2\Bigr]\psi_{\bm p}(t)=0,
\label{a10}
\end{equation}
where $\omega_{\bm p}=\sqrt{c^2{\bm p}^2+(m_{\rm e}c^2)^2}$. This harmonic oscillator equation has two linearly independent solutions, corresponding to energy 
$\omega_{\bm p}$ and $-\omega_{\bm p}$. In what follows, we will label these solutions with superscripts $\beta=+$ and $\beta=-$, respectively.
Namely, $\psi_{\bm p}^{(\beta)}(t)$ will be the solution of~\eqref{a9} which asymptotically, i.e., according to~\eqref{a10}, behaves as 
\begin{equation}
\psi_{\bm p}^{(\beta)}(t)\underset{t\rightarrow -\infty}{\sim}\ee^{-\ii\beta\omega_{\bm p}t}.
\label{a11}
\end{equation}
We will interpret these solutions as describing an electron ($\beta=+$) and its anti-particle, i.e., a positron ($\beta=-$) in a pulsed electric field. 
One can also show using Eq.~\eqref{a9} that under the charge conjugation and parity transformations (CP: $e\rightarrow-e$ and ${\bm p}\rightarrow-{\bm p}$),
\begin{equation}
\psi_{\bm p}^{(-)}(t)\rightarrow[\psi_{-{\bm p}}^{(+)}(t)]^*.
\label{a11new}
\end{equation}
Finally, the corresponding solutions of~\eqref{t3} have the form,
\begin{equation}
\chi_{{\bm p}\lambda}^{(\beta)}(x)=\ee^{\ii {\bm p}\cdot {\bm x}}\psi_{\bm p}^{(\beta)}(t)u_\lambda,
\label{a12}
\end{equation}
while those of the Dirac equation are obtained according to
\begin{align}
\Psi_{{\bm p}\lambda}^{(\beta)}(x)=\frac{1}{c}\Bigl[\ii\gamma^0\frac{\partial}{\partial t}&-c{\bm p}\cdot{\bm \gamma}+ceA(t)\gamma^3\nonumber\\
&+m_{\rm e}c^2\Bigr]\ee^{\ii {\bm p}\cdot {\bm x}}\psi_{\bm p}^{(\beta)}(t)u_\lambda.
\label{a13}
\end{align}
Note that asymptotically, for $t\rightarrow-\infty$, Eq.~\eqref{a13} is a linear combination of either free-particle or free-antiparticle solutions of the Dirac equation,
depending on the parameter $\beta$. Therefore, while asymptotically Eq.~\eqref{a13} describes an electron or a positron with momentum ${\bm p}$, these particles are 
in a superposition of spin up and down states.

As we show next, the just constructed eigenstates of the Dirac equation describing an electron/positron in a time-dependent electric field [Eq.~\eqref{a13}]
form a complete and orthonormal set of solutions~\cite{Grib1,Grib2}. One can check by direct calculations that
\begin{widetext}
\begin{align}
[\Psi_{{\bm p}\lambda}^{(\beta)}(x)]^\dagger \Psi_{{\bm p}\lambda'}^{(\beta')}(x)&=\frac{\delta_{\lambda\lambda'}}{c^2}\Bigl\{[\dot{\psi}_{\bm p}^{(\beta)}(t)]^*
\dot{\psi}_{\bm p}^{(\beta')}(t)+\ii c(p_{\|}-eA(t))\Bigl([\dot{\psi}_{\bm p}^{(\beta)}(t)]^*\psi_{\bm p}^{(\beta')}(t)-[\psi_{\bm p}^{(\beta)}(t)]^*
\dot{\psi}_{\bm p}^{(\beta')}(t)\Bigr)\nonumber\\
&+\omega_{\bm p}^2(t)[\psi_{\bm p}^{(\beta)}(t)]^*\psi_{\bm p}^{(\beta')}(t)\Bigr\},
\label{t10}
\end{align}
\end{widetext}
where the dot denotes the time derivative. This, in turn, allows us to prove that
\begin{equation}
\frac{\dd}{\dd t}\Bigl([\Psi_{{\bm p}\lambda}^{(\beta)}(x)]^\dagger\Psi_{{\bm p}\lambda'}^{(\beta')}(x)\Bigr)=0.
\label{t11}
\end{equation}
Thus, the quantity $[\Psi_{{\bm p}\lambda}^{(\beta)}(x)]^\dagger\Psi_{{\bm p}\lambda'}^{(\beta')}(x)$ 
is conserved during the time evolution. Specifically, using Eqs.~\eqref{a11} and~\eqref{t10}, one can derive that
\begin{equation}
\underset{t\rightarrow-\infty}{\lim}[\Psi_{{\bm p}\lambda}^{(\beta)}(x)]^\dagger\Psi_{{\bm p}\lambda'}^{(\beta')}(x)
=\frac{2\omega_{\bm p}}{c^2}(\omega_{\bm p}-\beta cp_{\|})\delta_{\lambda\lambda'}\delta_{\beta\beta'}.
\label{a15}
\end{equation}
This shows that the bispinors $\Psi_{{\bm p}\lambda}^{(\beta)}(x)$ can be normalized and, hence, we shall assume that
\begin{equation}
[\Psi_{{\bm p}\lambda}^{(\beta)}(x)]^\dagger\Psi_{{\bm p}\lambda'}^{(\beta')}(x)=\delta_{\lambda\lambda'}\delta_{\beta\beta'}.
\label{a16}
\end{equation}
Going further, the normalization condition for the eigenstates of the Dirac equation~\eqref{t1} takes the form,
\begin{equation}
\int\dd^3{\bm x}\,[\Psi_{{\bm p}\lambda}^{(\beta)}(x)]^\dagger\Psi_{{\bm p}'\lambda'}^{(\beta')}(x)=(2\pi)^3\delta({\bm p}-{\bm p}')\delta_{\lambda\lambda'}\delta_{\beta\beta'}.
\label{a17}
\end{equation}
Hence, the completeness relation for these eigenstates is
\begin{equation}
\sum_{\lambda=\pm}\sum_{\beta=\pm}\int\frac{\dd^3{\bm p}}{(2\pi)^3}\,\Psi_{{\bm p}\lambda}^{(\beta)}(x)[\Psi_{{\bm p}\lambda}^{(\beta)}(x')]^\dagger=\delta({\bm x}-{\bm x}').
\label{a18}
\end{equation}
The aforementioned analysis shows that the bispinors $\Psi_{{\bm p}\lambda}^{(\beta)}(x)$ form a complete set of orthonormal solutions of the Dirac equation in a pulsed
time-dependent electric field~\eqref{t1}~\cite{Grib1,Grib2}. These single particle solutions can be used now to construct the Dirac fermion field operator in the second quantization.

\subsection{Electron-positron pair creation from vacuum\\ by a time-dependent electric field}

The Dirac fermion field operator $\hat{\Psi}(x)$ is given by
\begin{equation}
\hat{\Psi}(x)=\sum_{\lambda}\int\frac{\dd^3 {\bm p}}{(2\pi)^3}\Bigl(\Psi_{{\bm p}\lambda}^{(+)}(x)\op{b}_{{\bm p}\lambda}+\Psi_{-{\bm p}\lambda}^{(-)}(x)\op{d}_{{\bm p}\lambda}^\dagger\Bigr),
\label{a19}
\end{equation}
where $\Psi^{(\beta)}_{\bm p}(x)$ are the one particle solutions of the Dirac equation~\eqref{a13}, whereas $\op{b}_{{\bm p}\lambda}$ and $\op{d}_{{\bm p}\lambda}$ are the annihilation operators of electron
and positron, respectively, in the eigenmode ${\bm p}\lambda$. These operators define the vacuum state at $t\rightarrow -\infty$ through the conditions
that $\op{b}_{{\bm p}\lambda}\ket{0_{-\infty}}=0$ and $\op{d}_{-{\bm p}\lambda}\ket{0_{-\infty}}=0$. Moreover, they satisfy the standard fermonic anti-commutation relations,
\begin{equation}
[\hat{b}_{{\bm p}\lambda},\op{b}_{{\bm p}'\lambda'}^\dagger]_+=[\hat{d}_{{\bm p}\lambda},\op{d}_{{\bm p}'\lambda'}^\dagger]_+=\delta({\bm p}-{\bm p}')\delta_{\lambda\lambda'},
\label{a20}
\end{equation}
with the remaining anti-commutators being zero. Keeping this in mind, we derive the instantaneous Hamiltonian of the fermion field $\op{H}(t)$~\cite{Grib1,Grib2},
\begin{widetext}
\begin{align}
\op{H}(t)=\sum_{\lambda}\int\frac{\dd^3{\bm p}}{(2\pi)^3}\Bigl[\gamma^{(++)}_{\bm p}(t)\op{b}_{{\bm p}\lambda}^\dagger\op{b}_{{\bm p}\lambda}+
\gamma^{(+-)}_{\bm p}(t)\op{b}_{{\bm p}\lambda}^\dagger\op{d}_{-{\bm p}\lambda}^\dagger
+\gamma^{(-+)}_{\bm p}(t)\op{d}_{-{\bm p}\lambda}\op{b}_{{\bm p}\lambda}+\gamma^{(--)}_{\bm p}(t)\op{d}_{-{\bm p}\lambda}\op{d}_{-{\bm p}\lambda}^\dagger\Bigr],
\label{a23}
\end{align}
where the coefficients $\gamma_{\bm p}^{(\beta\beta')}(t)$ are expressed as
\begin{equation}
\gamma_{\bm p}^{(\beta\beta')}(t)=\left\{ 
\begin{array}{ll}
-c(p_{\|}-eA(t))-\displaystyle\frac{2\epsilon_\perp^2}{c^2}{\rm Im}\Bigl([\psi_{\bm p}^{(\beta)}(t)]^*\dot{\psi}_{\bm p}^{(\beta)}(t)\Bigr) &\quad{\mbox{if}}\,\,\beta=\beta',\\
\displaystyle\frac{\ii\epsilon_\perp^2}{c^2}\Bigl([\psi_{\bm p}^{(\beta)}(t)]^*\dot{\psi}_{\bm p}^{(\beta')}(t)-[\dot{\psi}_{\bm p}^{(\beta)}(t)]^*\psi_{\bm p}^{(\beta')}(t)\Bigr)&\quad{\mbox{if}}\,\,\beta\neq\beta',
\end{array}
\right.
\label{a24}
\end{equation}
\end{widetext}
with $\epsilon_\perp=\sqrt{(c{\bm p}_\perp)^2+(m_{\rm e}c^2)^2}$.
With these definitions, one can verify that $\underset{t\rightarrow-\infty}\lim\gamma^{(\beta\beta')}(t)=\beta\omega_{\bm p}\delta_{\beta\beta'}$,
meaning that in the remote past the Hamiltonian~\eqref{a23} is diagonal. It becomes nondiagonal due to the interaction with the electric field, which is manifested
by the nonvanishing terms with $\op{b}_{{\bm p}\lambda}^\dagger\op{d}_{-{\bm p}\lambda}^\dagger$ and $\op{d}_{-{\bm p}\lambda}\op{b}_{{\bm p}\lambda}$. This affects
the vacuum state which becomes unstable.

In order to trace the vacuum instability, which results in pair creation, we introduce the Bogolyubov transformation~\cite{Bogolyubov},
\begin{align}
\op{b}_{{\bm p}\lambda}(t)&=\eta_{\bm p}(t)\op{b}_{{\bm p}\lambda}+\xi_{\bm p}(t)\op{d}^\dagger_{-{\bm p}\lambda},\label{a26a}\\
\op{d}_{{\bm p}\lambda}(t)&=\eta_{-{\bm p}}(t)\op{d}_{{\bm p}\lambda}-\xi_{-{\bm p}}(t)\op{b}_{-{\bm p}\lambda}^\dagger.
\label{a26b}
\end{align}
It introduces a new set of annihilation and, respectively, creation operators of quasiparticles at time $t$, such that the Hamiltonian is diagonal in the new
annihilation and creation operators. Hence, the instantaneous vacuum state is defined
as $\op{b}_{{\bm p}\lambda}(t)\ket{0_t}=0$ and $\op{d}_{{\bm p}\lambda}(t)\ket{0_t}=0$. Note that this transformation preserves the anti-commutation relations 
of the creation and annihilation operators provided that, at every time $t$, unknown functions $\eta_{\bm p}(t)$ and $\xi_{\bm p}(t)$ satisfy the condition,
\begin{equation}
|\eta_{\bm p}(t)|^2+|\xi_{\bm p}(t)|^2=1.
\label{a27}
\end{equation}
Hence, the temporal probability for a pair to be created in the state determined by ${\bm p}$ and $\lambda$ can be also defined~\cite{Grib1,Grib2}.
Namely, since for fermions no more than one pair can be created such that the electron carries the momentum ${\bm p}$ while the positron the momentum
$-{\bm p}$,
\begin{align}
{\cal P}(t)&=\bra{0_{-\infty}}\op{b}_{{\bm p}\lambda}^\dagger(t)\op{b}_{{\bm p}\lambda}(t)\ket{0_{-\infty}}\nonumber\\
&=\bra{0_{-\infty}}\op{d}_{-{\bm p}\lambda}^\dagger(t)\op{d}_{-{\bm p}\lambda}(t)\ket{0_{-\infty}}=|\xi_{\bm p}(t)|^2
\label{a28}
\end{align}
defines the aforementioned probability of pair creation at time $t$. In the following, we will be interested in the limit of~\eqref{a28} when $t\rightarrow +\infty$. 
For this purpose, one has to calculate the time-dependent coefficients of the Bogolyubov transformation first.

To this end, we rewrite the field operator~\eqref{a19} as
\begin{equation}
\op{\Psi}(x)=\sum_{\lambda}\int\frac{\dd^3{\bm p}}{(2\pi)^3}\Bigl[\Phi_{{\bm p}\lambda}^{(+)}(x)\op{b}_{{\bm p}\lambda}(t)+\Phi_{-{\bm p}\lambda}^{(-)}(x)
\op{d}_{{\bm p}\lambda}^\dagger(t)\Bigr],
\label{a29}
\end{equation}
with the bispinors $\Phi_{{\bm p}\lambda}^{(\beta)}(x)$ such that
\begin{align}
\Phi_{{\bm p}\lambda}^{(+)}(x)&=\eta_{\bm p}^*(t)\Psi_{{\bm p}\lambda}^{(+)}(x)+\xi_{\bm p}^*(t)\Psi_{{\bm p}\lambda}^{(-)}(x), \label{a30}\\
\Phi_{{\bm p}\lambda}^{(-)}(x)&=\eta_{\bm p}(t)\Psi_{{\bm p}\lambda}^{(-)}(x)-\xi_{\bm p}(t)\Psi_{{\bm p}\lambda}^{(+)}(x).\label{a31}
\end{align}
It follows from here that $\Phi_{{\bm p}\lambda}^{(\beta)}(x)$ should have the same spinor form as $\Psi_{{\bm p}\lambda}^{(\beta)}(x)$. Thus, we propose that
\begin{align}
\Phi_{{\bm p}\lambda}^{(\beta)}(x)&=\frac{1}{c}\Bigl[\ii\gamma^0\frac{\partial}{\partial t}-c{\bm p}\cdot{\bm \gamma}+ceA(t)\gamma^3\nonumber\\
&+m_{\rm e}c^2\Bigr]\ee^{\ii {\bm p}\cdot {\bm x}-\ii\beta\int^t\dd t'\omega_{\bm p}(t')}\phi_{\bm p}^{(\beta)}(t)u_\lambda,
\label{a32}
\end{align}
where $\phi_{\bm p}^{(\beta)}(t)$ are unknown functions. Now, combining Eqs.~\eqref{a13},~\eqref{a30},~\eqref{a31}, and~\eqref{a32}, we obtain that
\begin{align}
\psi_{\bm p}^{(+)}(t)=\eta_{\bm p}(t)&\ee^{-\ii\int^t\dd t'\omega_{\bm p}(t')}\phi_{\bm p}^{(+)}(t)\nonumber\\
&-\xi_{\bm p}^*(t)\ee^{\ii\int^t\dd t'\omega_{\bm p}(t')}\phi_{\bm p}^{(-)}(t),\label{a33}\\
\psi_{\bm p}^{(-)}(t)=\xi_{\bm p}(t)&\ee^{-\ii\int^t\dd t'\omega_{\bm p}(t')}\phi_{\bm p}^{(+)}(t)\nonumber\\
&+\eta_{\bm p}^*(t)\ee^{\ii\int^t\dd t'\omega_{\bm p}(t')}\phi_{\bm p}^{(-)}(t).\label{a34}
\end{align}
These functions satisfy Eq.~\eqref{a9} provided that
\begin{equation}
\phi_{\bm p}^{(\beta)}(t)=\frac{c}{\sqrt{2\omega_{\bm p}(t)\bigl[\omega_{\bm p}(t)-\beta c(p_{\|}-eA(t))\bigr]}}
\label{a35}
\end{equation}
and the coefficients $\eta_{\bm p}(t)$ and $\xi_{\bm p}(t)$ are coupled through equations,
\begin{align}
\dot{\eta}_{\bm p}(t)&=-\frac{ce{\cal E}(t)\epsilon_\perp}{2\omega_{\bm p}^2(t)}\,\xi_{\bm p}^*(t)\,\ee^{2\ii\int^t\dd t'\omega_{\bm p}(t')},\nonumber\\
\dot{\xi}_{\bm p}^*(t)& =\frac{ce{\cal E}(t)\epsilon_\perp}{2\omega_{\bm p}^2(t)}\,\eta_{\bm p}(t)\,\ee^{-2\ii\int^t\dd t'\omega_{\bm p}(t')}.
\label{a36}
\end{align}
Thus, we need to solve these equations numerically.

Before we proceed with calculations, let us note that in order for the functions~\eqref{a33} and~\eqref{a34}
to fulfill the condition~\eqref{a11new}, it must hold that under the CP transformation,
\begin{equation}
\eta_{-{\bm p}}(t)\rightarrow\eta_{\bm p}(t),\quad \xi_{-{\bm p}}(t)\rightarrow-\xi_{\bm p}(t).
\label{a37}
\end{equation}
This is in agreement with Eqs.~\eqref{a36}. Actually, the same can be figured out when imposing the CP transformation requirement on the field operator $\op{\Psi}(x)$~\cite{Berestetskii}. 
The point being that, in the second quantization, the CP transformation of $\op{\Psi}(x)$ can be formulated as transformation rules for the particle creation 
and annihilation operators. These rules imposed on the operators $\op{b}_{{\bm p}\lambda}(t)$ and $\op{d}_{{\bm p}\lambda}(t)$ lead to Eq.~\eqref{a37}.

As we have mentioned before, the Bogolyubov transformation allows one to diagonalize the Hamiltonian~\eqref{a23}. As we have checked this, it becomes
\begin{equation}
\op{H}(t)=\sum_{\lambda}\int\frac{\dd^3{\bm p}}{(2\pi)^3}\,\omega_{\bm p}(t)\Bigl[\op{b}_{{\bm p}\lambda}^\dagger(t)\op{b}_{{\bm p}\lambda}(t)
+\op{d}_{-{\bm p}\lambda}^\dagger(t)\op{d}_{-{\bm p}\lambda}(t)\Bigr],
\label{a42}
\end{equation}
where $\omega_{\bm p}(t)$ has the meaning of the instantaneous energy in the ${\bm p}\lambda$ eigenmode of the fermionic field $\op{\Psi}(x)$. 
Here, we have treated an infinite constant by means of the normal ordering of the creation and annihilation operators.

Actually, the system of equations~\eqref{a36} is not convenient for numerical analysis, as for the considered electric field strengths the phase factors 
$\ee^{\pm 2\ii\int^t\dd t'\omega_{\bm p}(t')}$ oscillate rapidly. For this reason, we introduce a new set of coefficients, $c_{\bm p}^{(1)}(t)$ and $c_{\bm p}^{(2)}(t)$, such that
\begin{align}
c_{\bm p}^{(1)}(t)&=\eta_{\bm p}(t)\,\ee^{-\ii\int^t\dd t'\omega_{\bm p}(t')},\label{a38}\\
c_{\bm p}^{(2)}(t)&=\xi_{\bm p}(t)\,\ee^{\ii\int^t\dd t'\omega_{\bm p}(t')}.\label{a39}
\end{align}
Then, Eq.~\eqref{a36} can be rewritten in the form~\cite{Akkermans},
\begin{equation}
\ii\frac{\dd}{\dd t}\begin{bmatrix}
                     c_{\bm p}^{(1)}(t)\\
                     c_{\bm p}^{(2)}(t)
                    \end{bmatrix}
                    =
                    \begin{pmatrix} \omega_{\bm p}(t) & \ii\Omega_{\bm p}(t) \cr -\ii\Omega_{\bm p}(t) & -\omega_{\bm p}(t) \end{pmatrix}
                    \begin{bmatrix}
                     c_{\bm p}^{(1)}(t)\\
                     c_{\bm p}^{(2)}(t)
                    \end{bmatrix},
\label{a40}
\end{equation}
with the off-diagonal matrix elements given by $\displaystyle\Omega_{\bm p}(t)=\frac{c|e|{\cal E}(t)\epsilon_\perp}{2\omega_{\bm p}^2(t)}$. The last
equation is structurally identical to the Schr\"odinger equation of a two-level system which undergoes a unitary time-evolution~\cite{Akkermans}. It will be solved
for an electric field model consisting of $N_{\rm rep}$ pulses (for more details, see Sec.~\ref{electric}). Hence, the probability of pair production 
from the vacuum by a sequence of $N_{\rm rep}$ electric field pulses into the eigenmode ${\bm p}\lambda$ is
\begin{equation}
{\cal P}_{N_{\rm rep}}=\lim_{t\rightarrow+\infty} |\xi_{\bm p}(t)|^2=\lim_{t\rightarrow+\infty}|c_{\bm p}^{(2)}(t)|^2,
\label{a41}
\end{equation}
which follows from Eqs.~\eqref{a28} and~\eqref{a39}.

In closing this section, let us comment on another approach which is widely used in this context and it is based on solving the quantum Vlasov equation (QVE)~\cite{Schmidt}
(see, Appendix~\ref{Vlasov}). While in our case, the system of differential equations~\eqref{a40} defines the
temporal probability amplitude of pair production, the quantum Vlasov equation is an integro-differential equation for the temporal probability 
of pair creation~\eqref{vla12}. In light of the results presented in Sec.~\ref{distributions}, one may ask whether interference patterns can be observed when solving the QVE. 
Since the QVE is a non-Markovian equation, the time-evolution of the respective probability
depends on the history of the fermionic field eigenmode interacting with the electric field, which is a memory effect. The physical importance of the memory is that it carries the information 
about quantum interference patterns, which has been confirmed in~\cite{Li2}.

\subsection{Electric field model}
\label{electric}

Similar to Akkermans and Dunne~\cite{Akkermans}, we consider the time-dependent electric field described by the shape function,
\begin{equation}
F_B(t)=\frac{1}{\cosh^2(t/\sigma)},
\label{sf1}
\end{equation}
with a free parameter $\sigma$. In contrast to their work, however, we will exclusively study the pair creation by electric field pulses, satisfying Eqs.~\eqref{a1} and~\eqref{a2}. 
For this reason, we assume in the following that a single electric field pulse is described by the shape function,
\begin{equation}
F_0(t)=N_0[F_B(t-T_0/2)-F_B(t+T_0/2)],
\label{sf3}
\end{equation}
where $F_B(t)$ is given by Eq.~\eqref{sf1}. Here, $T_0$ denotes the time-delay between both half-pulses, which is introduced 
in relation to the parameter $T$ present in~\cite{Akkermans}. If $T_0$ is sufficiently large ($T_0\gg\sigma$), in which case both half-pulses are well separated, 
then $N_0=1$. Otherwise, the constant $N_0$ is chosen such that
\begin{equation}
\mathrm{max}|F_0(t)|=1.
\label{sf4}
\end{equation}
Later on, we will compare the yield of created $e^-e^+$ pairs when induced by a single pulse~\eqref{sf3} and a finite train of such pulses.
The latter is defined by the shape function,
\begin{equation}
F(t)=\sum_{N=1}^{N_{\mathrm{rep}}}F_0\bigl[t+\bigl(2N-1-N_{\mathrm{rep}}\bigr)T/2\bigr],
\label{sf5}
\end{equation}
representing a sequence of $N_{\rm rep}$ identical copies of~\eqref{sf3}. Here, $T$ is chosen such that, within the numerical accuracy,
\begin{equation}
F_0(\pm T/2) = 0,
\label{sf6}
\end{equation}
which guarantees that the train consists of well-separated pulses. Thus, it has a clear physical meaning as a time-delay between the subsequent pulses.

This shape function defines the time-dependent electric field $\mathcal{E}(t)$ of the amplitude $\mathcal{E}_0$,
\begin{equation}
\mathcal{E}(t)=\mathcal{E}_0 F(t),
\label{sf7}
\end{equation}
and the corresponding time-dependent vector potential
\begin{equation}
A(t)=-\int_{-\infty}^t \mathcal{E}(\tau)\dd\tau=\int^{\infty}_t \mathcal{E}(\tau)\dd\tau .
\label{sf8}
\end{equation}
For $T$ such that the condition~\eqref{sf6} is satisfied, both the vector potential and the electric field vanish not only at infinities, 
but also at times in-between the pulses, i.e., for $t=(2N-N_{\mathrm{rep}})T/2$, where $N=1,\dots,N_{\mathrm{rep}}-1$. 
Keeping this in mind, we introduce the basic shape function characterizing the vector potential $f_B(t)$ such that it vanishes for $t\rightarrow +\infty$,
\begin{equation}
f_B(t)=\int^{\infty}_t F_B(\tau)\dd\tau .
\label{sf9}
\end{equation}
Namely,
\begin{equation}
f_B(t)=\sigma [1-\tanh(t/\sigma)],
\label{sf13}
\end{equation}
where we have used $F_B(t)$ given by~\eqref{sf1}. In addition, we define the vector potential shape function for the single pulse,
\begin{equation}
f_0(t)=\int_t^{\infty} F_0(\tau)\dd\tau=N_0[f_B(t+T_0/2)-f_B(t-T_0/2)] ,
\label{sf10}
\end{equation}
and for the train of pulses,
\begin{equation}
f(t)=\sum_{N=1}^{N_{\mathrm{rep}}}f_0\bigl[t+\bigl(2N-1-N_{\mathrm{rep}}\bigr)T/2\bigr].
\label{sf11}
\end{equation}
Hence,
\begin{equation}
A(t)=\mathcal{E}_0 f(t)
\label{sf12}
\end{equation}
is the vector potential describing physical pulses~\eqref{a1}.

\section{Probability distributions}
\label{distributions}
\begin{figure}
\includegraphics[width=8cm]{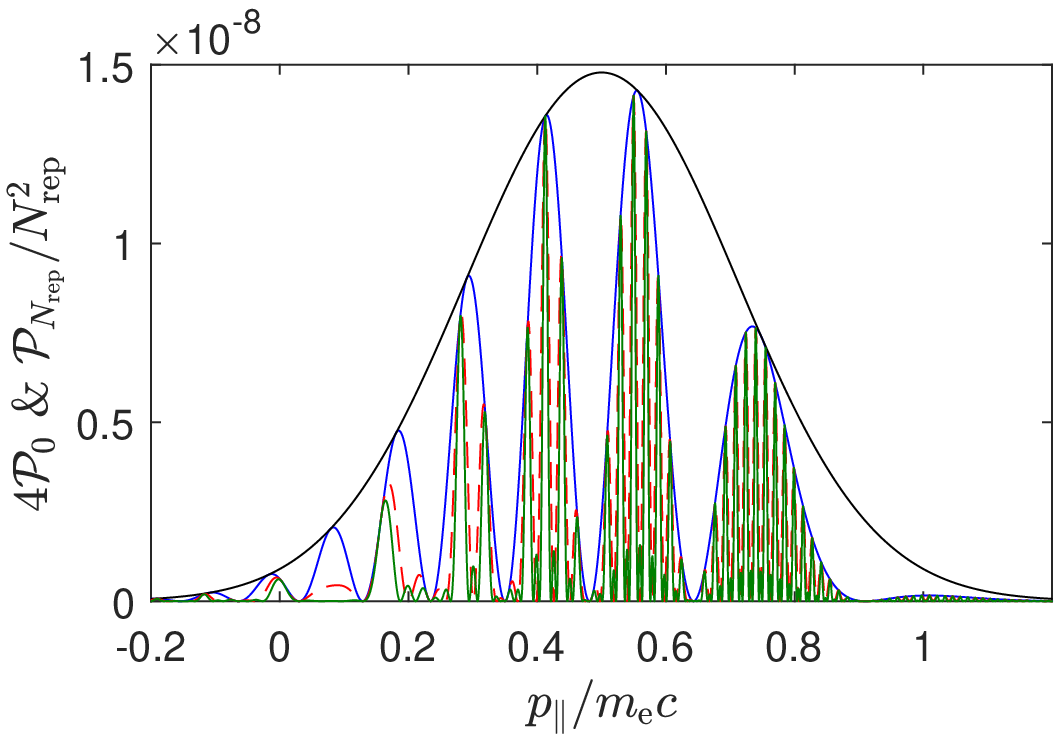}
\caption{Probability of pair creation from vacuum ${\cal P}_{N_{\rm rep}}$ as a function of the longitudinal momentum $p_{\|}$ (with ${\bm p}_\perp={\bm 0}$), when induced by a single pulse 
($N_{\mathrm{rep}}=1$) (solid blue line), or by a train of two ($N_{\mathrm{rep}}=2$) (dashed red line) or three pulses ($N_{\mathrm{rep}}=3$) (solid green line). 
The shape function of the driving electric field is defined by Eqs.~\eqref{sf1},~\eqref{sf3}, and~\eqref{sf5} with the following parameters: $\sigma=5\tau_{\rm C}$, 
$T_0=40\tau_{\rm C}$, and $T=400\tau_{\rm C}$, where $\tau_{\rm C}=1/(m_{\rm e}c^2)$ is the Compton time. The amplitude of the electric field ${\cal E}_0$ (in units of the 
Sauter-Schwinger critical field, ${\cal E}_{\rm S}=m_{\rm e}^2c^3/|e|$) is ${\cal E}_0=-0.1{\cal E}_{\rm S}$. The results are compared with the pair probability distribution
induced by a half-pulse ${\cal P}_0$ (solid black envelope), characterized by the shape function~\eqref{sf1}.
\label{rf2u1}}
\end{figure} 
\begin{figure}
\includegraphics[width=8cm]{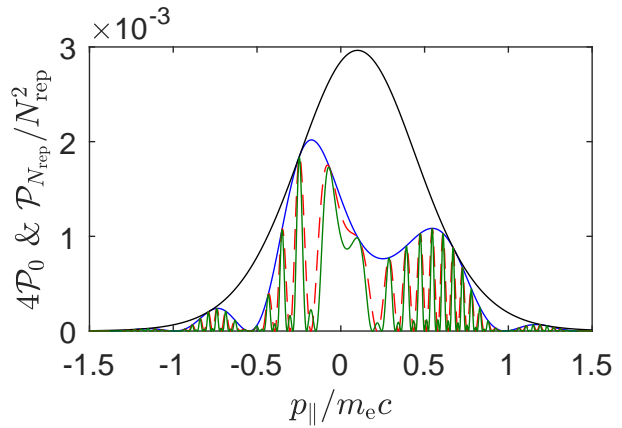}
\caption{The same as in Fig.~\ref{rf2u1} but for $\sigma=\tau_{\rm C}$, $T_0=10\tau_{\rm C}$, and $T=100\tau_{\rm C}$. 
\label{rf2w1}}
\end{figure}

In Fig.~\ref{rf2u1}, we plot the probability of pair creation from vacuum, ${\cal P}_{N_{\rm rep}}$, by a time-dependent pulsed electric field, which has been defined 
in Sec.~\ref{electric}. The distribution denoted by the solid blue line describes the process driven by a single pulse ($N_{\rm rep}=1$). The results plotted as the dashed red line
and the solid green line correspond to the pair creation driven by a train of pulses with either two ($N_{\rm rep}=2$) or three pulse repetitions ($N_{\rm rep}=3$), respectively.
These results are scaled by $N_{\rm rep}^2$. In addition, we show the probability distribution when generated by a half-pulse electric field~\eqref{sf1} 
(solid black envelope). Similar to~\cite{Akkermans}, this distribution is multiplied by a factor of four, i.e., it is scaled to match the maximum distribution for 
the single pulse ($N_{\rm rep}=1$). For the electric field parameters we keep $\sigma=5\tau_{\rm C}$, $T_0=40\tau_{\rm C}$, and $T=400\tau_{\rm C}$, which are expressed in units of 
the Compton time, $\tau_{\rm C}=1/(m_{\rm e}c^2)$. The field amplitude ${\cal E}_0$ (in units of the Sauter-Schwinger electric field, ${\cal E}_{\rm S}=m_{\rm e}^2c^3/|e|$)
is ${\cal E}_0=-0.1{\cal E}_{\rm S}$. The spectra are plotted as functions of the longitudinal momentum $p_{\|}$, i.e., for ${\bm p}_\perp={\bm 0}$. 
Such a choice is justified as the particles are mostly generated in the direction of the electric field oscillations. While the distribution obtained for the half-pulse~\eqref{sf1}
exhibits a broad structure, already for the single pulse~\eqref{sf3} we observe typical modulations in the spectrum of produced pairs. Such modulations have been seen in Ref.~\cite{Akkermans}
and attributed to a double-slit Ramsey interference in time-domain, with each half-pulse acting as a slit. Taking into account our definition of a pulsed electric field,
which satisfies the condition~\eqref{a2}, it is justified to refer to such pattern as caused by an {\it intra-pulse interference}.

When applying a sequence of pulses to the QED vacuum additional peak structures in the spectrum appear (see, the results for two and three pulse repetitions 
in Fig.~\ref{rf2u1}). Such structures are much finer than the intra-pulse modulations. Typically, they consist of maxima which appear at the same values
of the longitudinal momenta $p_{\|}$, independently of $N_{\rm rep}$. At these momenta, the probability distributions ${\cal P}_{N_{\rm rep}}$ approximately scale to the one resulting 
from a single pulse interaction with vacuum (solid blue curve), with a typical scaling factor $N_{\rm rep}^2$. In addition, 
these peaks become more narrow with increasing the number of pulses in the train, i.e., with increasing $N_{\rm rep}$. Note that the main peaks in the spectra 
are accompanied by secondary maxima. For a given $N_{\rm rep}$, there is always $(N_{\rm rep}-2)$ of such secondary peaks. Actually, all these features can be seen more 
easily in Fig.~\ref{rf2w1}. Here, the spectra are presented for the same parameters as in Fig.~\ref{rf2u1}, except that the time delays between half-pulses and 
consecutive pulses are smaller now, $T_0=10\tau_{\rm C}$ and $T=100\tau_{\rm C}$, respectively. Also, the half-pulse width is smaller, $\sigma=\tau_{\rm C}$. While modulations 
of the peak structures originate from the intra-pulse interference, the peaks themselves occur only when a sequence of electric field pulses [in the sense of Eq.~\eqref{a2}] 
is applied. Hence, we conclude that their origin must be due to {\it inter-pulse interferences}. Note that a distinction between different patterns in the 
probability distributions as being due to either intra- or inter-pulse interferences is possible because we study only those pulsed electric fields which satisfy the physical 
condition~\eqref{a2}.

Now, let us discuss properties of the spectra presented in Figs.~\ref{rf2u1} and~\ref{rf2w1} in relation to the electric field parameters.
First of all, already for half a pulse~\eqref{sf1}, one observes a significant (i.e., roughly five orders of magnitude) difference when comparing the spectra. 
While the electric field amplitude applied in both figures is the same, it has to be related to the parameter $\sigma$. We recall that $1/\sigma$
describes the bandwidth of the pulsed electric field~\eqref{sf1}, which is broader in the case considered in Fig.~\ref{rf2w1}. As a result, the electric field quanta 
of larger energies interact with the QED vacuum, making the process of pair creation more probable. Next, we analyze 
modulations of the pair momentum distribution, which are denoted in both figures by the solid blue curves. These modulations are slower in Fig.~\ref{rf2w1},
which is related to the shorter time delay $T_0$ between both half pulses driving the pair creation. Finally, a shorter delay between the consecutive
pulses from the train $T$ (Fig.~\ref{rf2w1}) makes for broader individual peaks in the momentum distributions for $N_{\rm rep}>1$ and increases their separation.
These fine properties of the momentum spectra (or, equivalently, of the energy spectra) of particles can be explained based on the time-energy uncertainty principle.
Assuming that $T_0$ and $T$ define characteristic times over which the energy (momentum) of the system changes rapidly, more abrupt changes should be
observed for longer times, which is the case considered in Fig.~\ref{rf2u1}.

While the above analysis proves the sensitivity of the resulting distributions to the external field parameters, the question arises: Under which conditions
do the peak structures in the momentum distributions of the created particles arise when the process is driven by a train of identical pulses? We answer this question 
next, when analyzing properties of the time evolution operator.

\section{Avoided crossings vs crossings\\ of the evolution matrix eigenvalues}
\label{evolution}

Formally, we have reduced the problem to investigating the dynamics of the two-level system, which is governed by Eq.~\eqref{a40}. Now, we will use this similarity
to interpret our numerical results presented in the previous section.

The time evolution of such a system is given by a unitary $2\times 2$ matrix $\op{U}(t,t')$, $t\geqslant t'$, that satisfies the equation,
\begin{equation}
\ii\frac{\dd}{\dd t}\op{U}(t,t')
=\begin{pmatrix} \omega_{\bm p}(t) & \ii\Omega_{\bm p}(t) \cr -\ii\Omega_{\bm p}(t) & -\omega_{\bm p}(t) \end{pmatrix}
\op{U}(t,t'),
\label{pro4}
\end{equation}
with the initial condition $\op{U}(t',t')=\op{I}$. For a train of $N_{\rm rep}$ identical pulses driving the pair creation, the functions $\omega_{\bm p}(t)$
and $\Omega_{\bm p}(t)$ are periodic in the interval $N_{\rm rep}T$, with a period $T$ defining the time duration of an individual pulse from the train. Thus,
the system dynamics is determined by its evolution over time $T$. Lets denote the respective time evolution operator as $\op{U}(T+t',t')\equiv\op{U}(T)$.
It follows from the composition condition,
\begin{equation}
\op{U}(t,t')=\op{U}(t,t'')\op{U}(t'',t'),
\label{composition}
\end{equation}
where $t''$ is an intermediate time between $t'$ and $t$, that 
\begin{equation}
\op{U}(t'+{N_\mathrm{rep}}T,t')=[\op{U}(T)]^{N_\mathrm{rep}}.
\label{composition_new} 
\end{equation}
Keeping this in mind, we introduce the eigenvalue problem for the operator $\op{U}(T)$ (which is also called the monodromy matrix~\cite{monodromy}),
\begin{equation}
\op{U}(T)\ket{j}=\ee^{-\ii\vartheta_j}\ket{j}, \quad j=1,2,
\label{pro5}
\end{equation}
where the eigenvalues, $\ee^{-\ii\vartheta_j}$, are chosen as complex numbers with the modulus equal to one, and $\ket{j}$ denote their corresponding eigenstates.
As discussed in Appendix~\ref{appendix1}, $\vartheta_j$ are defined modulo $2\pi$ and the eigenstates $\ket{j}$ can be parametrized as
\begin{align}
\ket{1}&=\ee^{\ii\psi_1}\begin{pmatrix} \ee^{-\ii\beta/2}\cos(\gamma/2) \cr
\ee^{\ii\beta/2}\sin(\gamma/2) \end{pmatrix}, \nonumber \\
\ket{2}&=\ee^{\ii\psi_2}\begin{pmatrix} -\ee^{-\ii\beta/2}\sin(\gamma/2) \cr
\ee^{\ii\beta/2}\cos(\gamma/2) \end{pmatrix} ,
\label{appa3}
\end{align}
where $0\leqslant\gamma\leqslant\pi$, $0\leqslant\beta<2\pi$, and the global phases $\psi_j$ can be chosen arbitrary as they are irrelevant in our further analysis.
We next use the spectral decomposition of $\op{U}(T)$,
\begin{equation}
\op{U}(T)=\ee^{-\ii\vartheta_1}\op{P}_1+\ee^{-\ii\vartheta_2}\op{P}_2,
\label{pro6}
\end{equation}
where $\op{P}_j=\ket{j}\bra{j}$ (for $j=1,2$) is the projection operator on the state $\ket{j}$. It follows from Eqs.~\eqref{composition_new} and~\eqref{pro6} that
\begin{widetext}
\begin{align}
[\op{U}(T)]^{N_\mathrm{rep}}=\ee^{-\ii {N_\mathrm{rep}}\vartheta_1}\op{P}_1+\ee^{-\ii {N_\mathrm{rep}}\vartheta_2}\op{P}_2
=\ee^{-\ii N_{\rm rep}\vartheta_0}
\begin{pmatrix}
\cos(N_{\rm rep}\vartheta)+\ii\sin(N_{\rm rep}\vartheta)\cos\gamma & \ii\ee^{-\ii\beta}\sin(N_{\rm rep}\vartheta)\sin\gamma
\cr
\ii\ee^{\ii\beta}\sin(N_{\rm rep}\vartheta)\sin\gamma & \cos(N_{\rm rep}\vartheta)-\ii\sin(N_{\rm rep}\vartheta)\cos\gamma
\end{pmatrix} ,
\label{pro7}
\end{align}
\end{widetext}
where we introduce $\vartheta_0=(\vartheta_2+\vartheta_1)/2$ and $\vartheta=(\vartheta_2-\vartheta_1)/2$. Thus, there are four real angles 
$0\leqslant \vartheta_0,\vartheta < 2\pi$, $\beta$, and $\gamma$ which define the evolution of the system while it interacts with the pulsed electric field.
As we show next, only two of them define the probability of pair creation.

The dynamics of each eigenmode of the fermionic field $\op{\Psi}(x)$, which belongs to the momentum ${\bm p}$, is governed by the time-dependent Hamiltonian,
\begin{equation}
\op{H}_{\bm p}(t)=\begin{pmatrix} \omega_{\bm p}(t) & \ii\Omega_{\bm p}(t) \cr -\ii\Omega_{\bm p}(t) & -\omega_{\bm p}(t) \end{pmatrix}.
\label{ham1}
\end{equation}
Note that, in the remote past and future, it becomes
\begin{equation}
\op{H}_{\bm p}=\lim_{t\rightarrow\pm\infty}\op{H}_{\bm p}(t)=\begin{pmatrix} \omega_{\bm p} & 0 \cr 0 & -\omega_{\bm p} \end{pmatrix}.
\label{ham2}
\end{equation}
This means that, asymptotically, each eigenmode of the fermionic field with the momentum ${\bm p}$ 
can have the energy $\omega_{\bm p}$ or $-\omega_{\bm p}$. Hence, we interpret the upper energy eigenstate $\ket{+}=(1,0)^T$ as the one that 
describes an electron, whereas the lower energy eigenstate $\ket{-}=(0,1)^T$ describes a positron (here, $T$ means the transposition). Once the electric 
field is turned on, it couples these eigenstates leading to creation of a pair. Namely, an electron occupying the lower energy level is promoted by the electric field to 
the higher energy level, which otherwise is vacant, and a real electron and a hole are being created. Such a transition, which is due to the interaction with $N_{\rm rep}$
identical electric field pulses, occurs with probability,
\begin{align}
{\cal P}_{N_{\rm rep}}=|\bra{+}[\op{U}(T)]^{N_{\rm rep}}\ket{-}|^2=\sin^2\gamma\sin^2(N_{\rm rep}\vartheta),
\label{ham3}
\end{align}
where we have used Eq.~\eqref{pro7}. For completeness, we also write down the probability of the pair creation by a single pulse ($N_{\rm rep}=1$), 
\begin{equation}
{\cal P}_1=|\bra{+}\op{U}(T)\ket{-}|^2=\sin^2\gamma\sin^2\vartheta.
\label{ham4} 
\end{equation}
These two equations allow one to interpret the modulations of probability distributions of pair creation, which have been presented in the previous section.

\begin{figure}
\includegraphics[width=7.5cm]{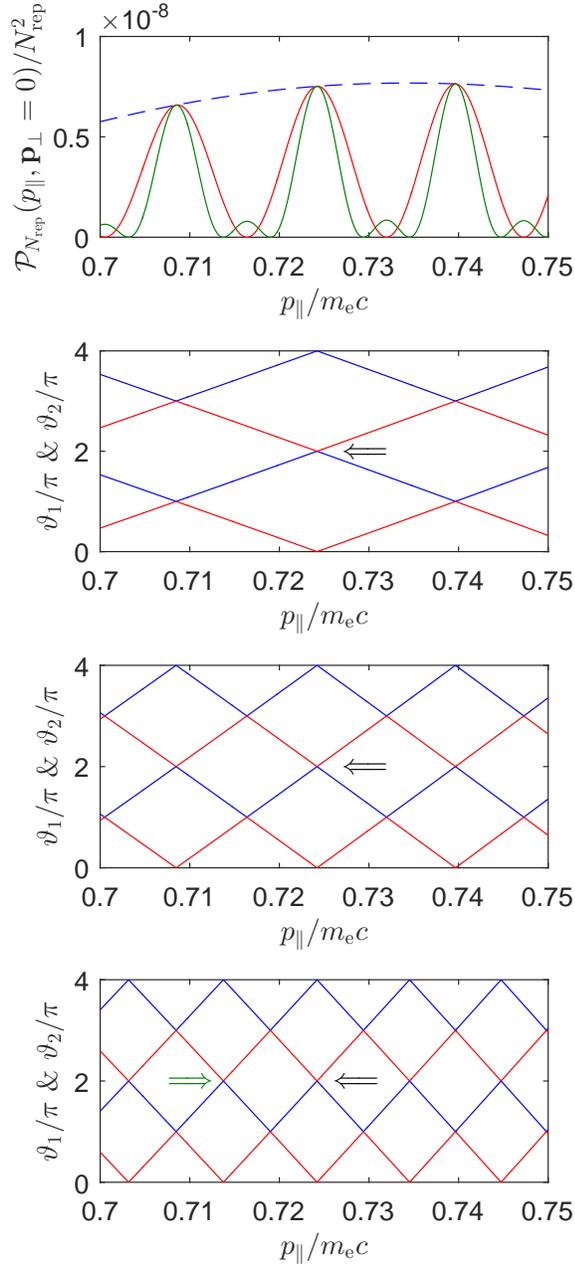}
\caption{The top panel shows the longitudinal momentum distribution of particles created in the process driven by a single electric field pulse
($N_{\mathrm{rep}}=1$) (dashed blue line), by a train of two such pulses ($N_{\mathrm{rep}}=2$) (solid red line), and by three such 
pulses ($N_{\mathrm{rep}}=3$) (solid green line). This panel relates to the same parameters of the electric field as Fig.~\ref{rf2u1} except that,
for visual purposes, we consider now a smaller range of the longitudinal momentum. Below we plot the phases $\vartheta_1\in(-\pi,0)$ (modulo $2\pi$) (in blue) 
and $\vartheta_2\in (0,\pi)$ (modulo $2\pi$) (in red) which define the eigenvalues of the time evolution operator~\eqref{pro7}, as functions of $p_{\|}$. 
Each consecutive panel (from top to bottom) corresponds to $N_{\rm rep}=1, 2$, and 3, respectively. Crossings 
and avoided crossings of the $\vartheta_1$ and $\vartheta_2$ curves correspond to either zeros of the probability distribution or to its maxima. 
A sample crossing and avoided crossing are indicated by the green and the black arrows, respectively.
\label{r20rfaz0d180216a}}
\end{figure}

\begin{figure}
\includegraphics[width=7.5cm]{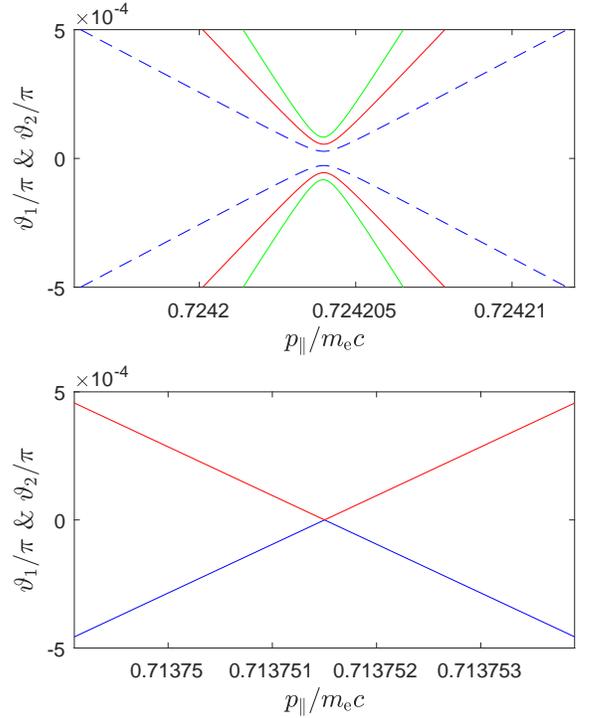}
\caption{Details of the avoided crossing and the crossing of the $\vartheta_1$ and $\vartheta_2$ curves indicated in Fig.~\ref{r20rfaz0d180216a}
by the black and the green arrows, respectively. In the upper panel, the results are for different $N_{\rm rep}$. Specifically, the dashed blue curve is
for a single electric field pulse ($N_{\rm rep}=1$), the solid red line is for a sequence of two such pulses ($N_{\rm rep}=2$), and the solid green
curve is for three such pulses ($N_{\rm rep}=3$). We see in this panel that the gap between both phases at the avoided crossing increases linearly with $N_{\rm rep}$. 
In the lower panel, the true crossing is presented for $N_{\rm rep}=3$.
\label{r20rfaz99d180216}}
\end{figure}

As it follows from Eqs.~\eqref{ham3} and~\eqref{ham4}, the probability of pair creation into the eigenmode with the momentum ${\bm p}$ depends on two 
angles parameterizing the time evolution matrix~\eqref{pro7}, $\gamma$ and $\vartheta$. The latter is related to the phases of complex eigenvalues of the
time evolution matrix~\eqref{pro6}, as $\vartheta=(\vartheta_2-\vartheta_1)/2$. If these phases are the same modulo $2\pi$, 
the probabilities of pair creation by a train of electric field pulses or by an individual electric field pulse are expected 
to be zero. This relates to the fact that, in such case, the time evolution of the system is trivial. Namely, it follows from Eq.~\eqref{pro6} that
$\op{U}(T)=\ee^{-\ii \vartheta_1}\op{I}$, which means that the lower and the higher energy eingenstates are uncoupled and the transition between them does not
occur. As discussed next, our numerical results confirm this expectation.

In the top panel of Fig.~\ref{r20rfaz0d180216a}, we show a portion of the probability distribution 
presented in Fig.~\ref{rf2u1}. This time, a dashed blue curve corresponds to the pair creation by a single electric field pulse ($N_{\rm rep}=1$),
the solid red curve is for the sequence of two electric field pulses ($N_{\rm rep}=2$), whereas the solid green curve is for three such pulses ($N_{\rm rep}=3$).
Below, in Fig.~\ref{r20rfaz0d180216a}, we present the dependence of the phases $\vartheta_1$ (in blue) and $\vartheta_2$ (in red) on the longitudinal momentum. 
For visual purposes, the phases have been defined such that $\vartheta_1\in (-\pi,0)$ (modulo $2\pi$) whereas $\vartheta_2\in (0,\pi)$ (modulo $2\pi$).
The subsequent panels correspond to $N_{\rm rep}=1, 2$, and 3. As confirmed by our numerical analysis, whenever these curves intersect the probability distribution is zero.
For example, one of such intersections indicated in the bottom panel by the green arrow (the one pointing to the right) can be traced to the zero of the respective probability distribution
for $N_{\rm rep}=3$ (green solid line) in the very top panel. It is also shown on the enlarged scale in the bottom panel of Fig.~\ref{r20rfaz99d180216}.
Note, however, that there are such points which seem to be the {\it crossings} of the $\vartheta_1$ and $\vartheta_2$ curves but actually they are {\it avoided crossings}. 
These are, for instance, all points which resemble the crossings in the second panel from the top in Fig.~\ref{r20rfaz0d180216a}, which is for $N_{\rm rep}=1$. 
Following the one marked there by the black arrow (pointing to the left), we see that it persists while increasing the number of pulses driving the pair creation 
(see, the remaining two panels of Fig.~\ref{r20rfaz0d180216a}). It is also presented on the enlarged scale in the upper panel of Fig.~\ref{r20rfaz99d180216} 
for $N_{\rm rep}=1,2$, and 3. Remarkably, very quick oscillations of the probability distributions observed for $N_{\rm rep}>1$ coincide with the regions where 
two phases $\vartheta_1$ and $\vartheta_2$ exhibit their avoided crossings. On the other hand, the same crossings occur already for $N_{\rm rep}=1$, which does not result
in sharp oscillations of the probability distributions of pair production. Thus, it must be the combined behavior of $\vartheta$ and $\gamma$ which determine 
the properties of the observed distributions, as we explain next.

\begin{figure}
\includegraphics[width=7.5cm]{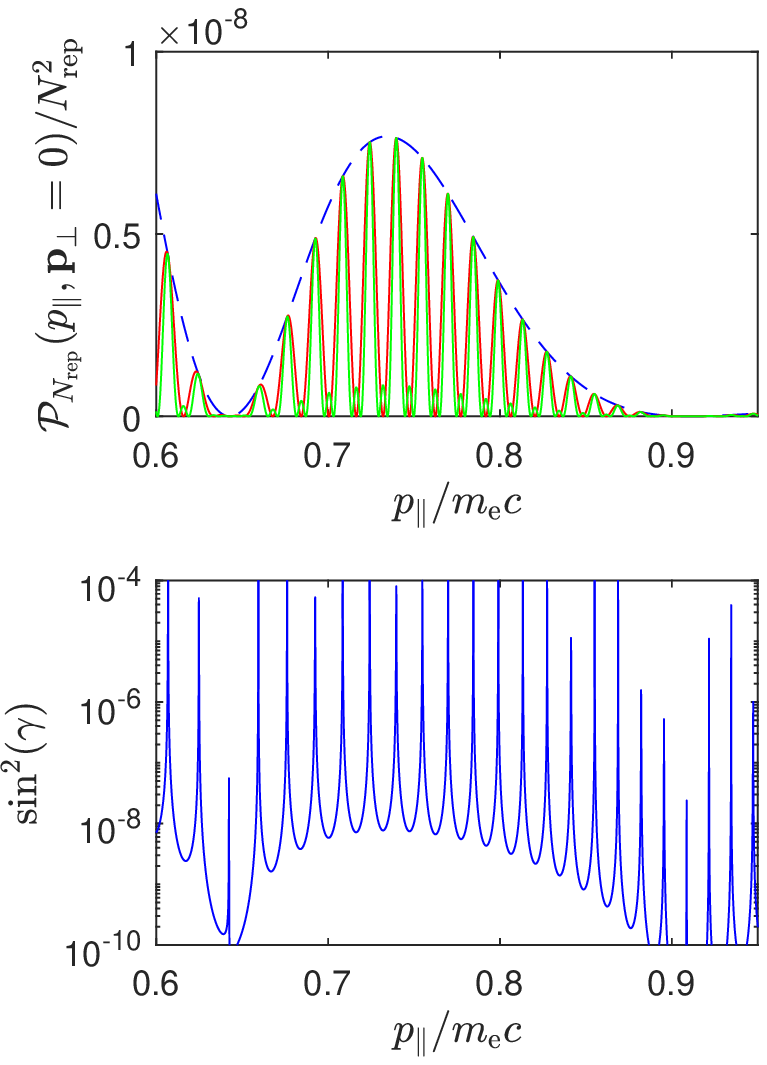}
\caption{Portion of the probability distributions of pair creation from Fig.~\ref{rf2u1} (upper panel) and $\sin^2(\gamma)$ (lower panel)
as functions of the longitudinal momentum $p_{\|}$ . The same color coding is used 
in the upper panel as in the top panel of Fig.~\ref{r20rfaz0d180216a}.
\label{r20rfaz1d180217}}
\end{figure}
\begin{figure}
\includegraphics[width=8cm]{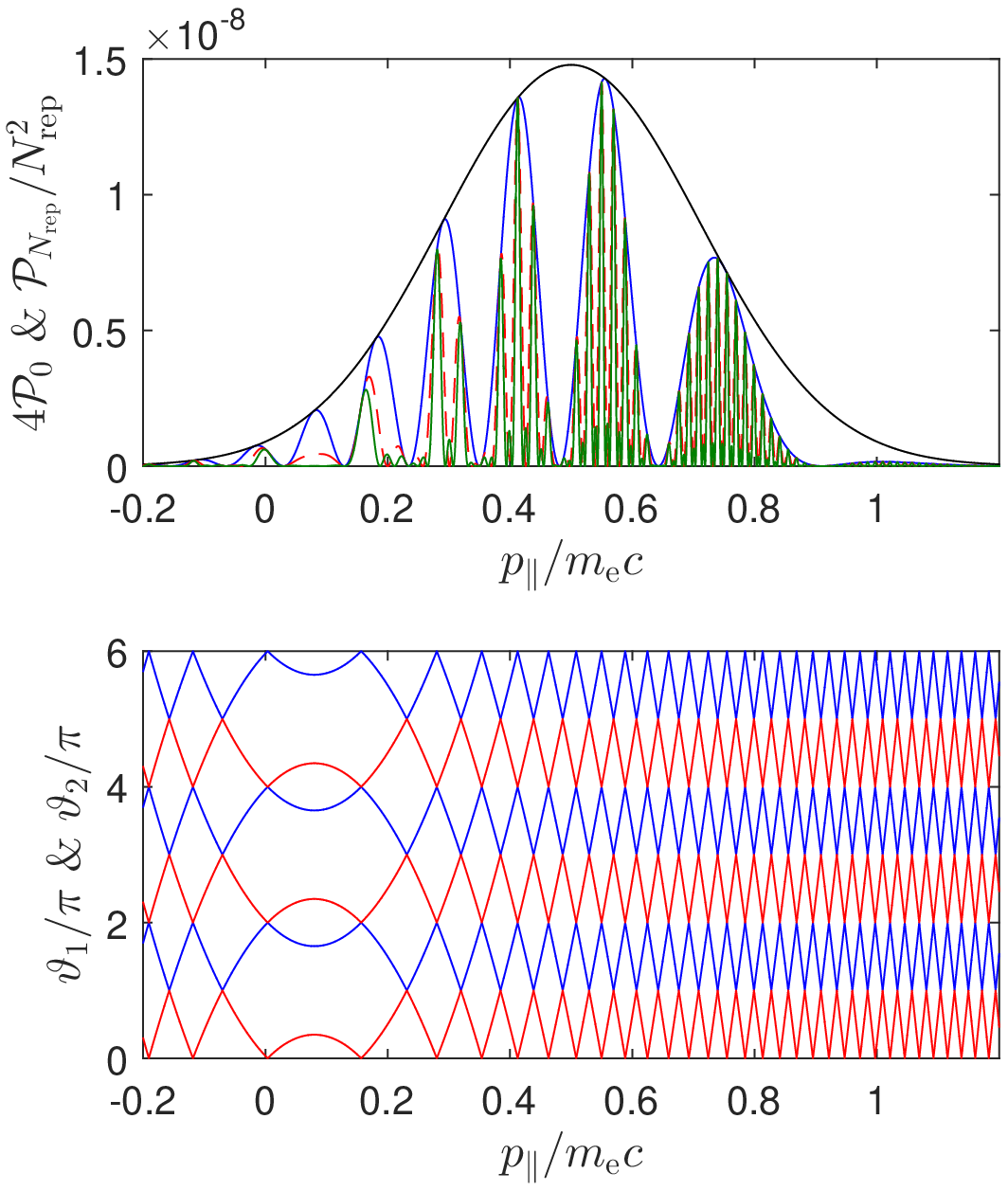}
\caption{The same as in Fig.~\ref{rf2u1} except that in the lower panel we plot the dependence of $\vartheta_1$ and $\vartheta_2$ [which parametrize the eigenvalues
of the evolution matrix, see, Eq.~\eqref{pro7}] on the longitudinal momentum for $N_{\rm rep}=1$. Although the probability distribution ${\cal P}_1$
is a smooth broad curve, the avoided crossings of the phases $\vartheta_1$ and $\vartheta_2$ for $N_{\rm rep}=1$ allow to predict at which momenta $p_{\|}$
the probability distribution of created particles will peak for a train of such pulses.
\label{rf2u1ex}}
\end{figure}

In the lower panel of Fig.~\ref{r20rfaz1d180217}, we plot the dependence of $\sin^2\gamma$ on the longitudinal momentum $p_{\|}$. Note that 
$\sin^2\gamma$ determines the probability of pair creation [see, Eq.~\eqref{ham3} and~\eqref{ham4}]. As one can see, $\sin^2\gamma$ takes in general rather 
small values. However, it varies abruptly around specific $p_{\|}$. As it follows from the upper panel of Fig.~\ref{r20rfaz1d180217}, at these values of $p_{\|}$ 
the probability distributions of pairs generated from the vacuum by a sequence of pulses ($N_{\rm rep}>1$) exhibit their local maxima. This indicates that 
resonant-like peaks of $\sin^2\gamma$ coincide with the avoided crossings of the phases $\vartheta_1$ and $\vartheta_2$. As a consequence, a smooth dependence 
of the probability distribution on $p_{\|}$ for $N_{\rm rep}=1$ is observed [Eq.~\eqref{ham4}]. For $N_{\rm rep}>1$, the situation is different. We rewrite 
Eq.~\eqref{ham4} such that
\begin{equation}
{\cal P}_{N_{\rm rep}}={\cal P}_1 \Bigl[\frac{\sin(N_{\rm rep}\vartheta)}{\sin\vartheta}\Bigr]^2,
\label{ham5}
\end{equation}
where the smooth function of $p_{\|}$, ${\cal P}_1$, is multiplied by the so-called diffraction term (also known as the interference term). This term arises in optics when considering diffraction 
of light by a grating of $N_{\rm rep}$ slits in the far-field zone. It also leads to a coherent $N_{\rm rep}^2$-type of enhancement of the light intensity at specific 
values of $\vartheta$. However, in our case, such perfect $N_{\rm rep}^2$ coherence is never achieved. The reason is that, while the diffraction factor
takes the maximum value for $\bar{\vartheta}_n=n\pi$ ($n=0,\pm 1,...$), which equals $N_{\rm rep}^2$, at those points ${\cal P}_1$ and ${\cal P}_{N_{\rm rep}}$ are zero. Instead,
we observe a nearly perfect enhancement of the pair production signal at $\bar{\vartheta}_n\approx \vartheta_{\rm G}+n\pi$ or, equivalently, at $\vartheta_2\approx\vartheta_1
+2\vartheta_{\rm G}$ (modulo $2\pi$), with $2\vartheta_{\rm G}$ being the gap between the two curves $\vartheta_1$ and $\vartheta_2$. At those points,
\begin{equation}
\Bigl[\frac{\sin({N_\mathrm{rep}}\vartheta)}{\sin\vartheta} \Bigr]^2\approx N_\mathrm{rep}^2\bigl[1-\frac{4}{3}(N_{\rm rep}^2-1)\vartheta_G^2\bigr],
\label{pro11}
\end{equation}
where one has to assume that $\vartheta_{\rm G}\ll 1$ and $N_{\rm rep}\vartheta_{\rm G}\ll 1$. If these conditions are not satisfied, i.e., the gap between the $\vartheta_1$
and $\vartheta_2$ curves becomes significant, the nearly perfect coherence is lost. This is confirmed by our numerical results.
Specifically, one can see that in the upper panel of Fig.~\ref{rf2u1ex} for a main maximum centered around $0.08m_{\rm e}c$. In this case, the results
corresponding to different $N_{\rm rep}$ do not scale according to~\eqref{pro11}. This is because the gap between the $\vartheta_1$ and $\vartheta_2$ curves,
as shown in the lower panel of Fig.~\ref{rf2u1ex}, is too large and Eq.~\eqref{pro11} does not apply.

In studying properties of the probability distributions shown in Sec.~\ref{distributions}, we observe that the fine peaks present in the spectra for $N_{\rm rep}>1$
are not equidistant. This is related to the positions of avoided crossings of the phases $\vartheta_1$ and $\vartheta_2$ as functions of the longitudinal momentum 
$p_{\|}$ and, as discussed above, is determined by their behavior already for $N_{\rm rep}=1$. Note that in Fig.~\ref{r20rfaz0d180216a}, the $\vartheta_1$ and 
$\vartheta_2$ curves occur as straight lines (for $N_{\rm rep}=1$)
with same slopes; thus, suggesting that avoided crossings are positioned at a fixed $p_{\|}$ increment. The reason is that the momentum region considered there 
is very small. In fact, if plotted over a larger interval of $p_{\|}$ (Fig.~\ref{rf2u1ex}), the phases turn out to be nearly parabolas for small $|p_{\|}|$ with increasingly steeper arms at larger $p_{\|}$.
This makes for a denser distribution of avoided crossings for larger momenta $p_{\|}$ and, hence, also for a denser distribution of the inter-pulse peaks in Fig.~\ref{rf2u1}. 
The same concerns Fig.~\ref{rf2w1}, but it is related to $|p_{\|}|$ instead.

Note that a very similar expression for the number of created pairs, i.e., with a diffraction-type factor, has been derived in~\cite{Akkermans}. Their analysis, however,
was based on the tunneling picture of pair creation which is appropriate for $\sigma\gg 1/(m_{\rm e}c^2)$. In such case, it is justified to use the WKB theory and the turning
points analysis in deriving the aforementioned formula. Our approach, on the other hand, is free of this assumption. Specifically, it also works very well for parameters
used in Fig.~\ref{rf2w1}, where the quasiclassical approximation cannot be applied. We also would like to mention that both approaches refer to different concepts of 
avoided crossings. In~\cite{Akkermans}, the avoided crossings are the turning points which are complex-time solutions of the equation $\omega_{\bm p}(t)=0$. It is the 
respective sum over all dominant turning points which results in a coherent enhancement of the number of created pairs. In our approach, the avoided 
crossings correspond to the real phases $\vartheta_1$ and $\vartheta_2$ of the eigenvalues of the time-evolution operator $\op{U}(T)$. They also introduce a dephasing
mechanism, which may lead to a significant loss of coherence for sufficiently strong electric field pulses. Despite these obvious differences, both approaches provide legitimate explanations for appearance of 
the peak structure in the signal of produced pairs, the origin of which lies in diffraction of the vacuum at a time grating formed by a sequence of time-dependent electric field pulses.

\section{Conclusions}
\label{conclusions}

In this paper, we have analyzed the electron-positron pair creation from vacuum driven by a sequence of $N_{\rm rep}$ identical electric-field pulses. 
We have shown that intra- and inter-pulse interference structures in the probability distributions of created particles arise in such scenario. 
Importantly, this is independent of the electric-field parameters. For instance, the same is observed beyond the regime of applicability of the WKB theory which has been
the focus of earlier studies~\cite{Akkermans,Li1,Li2}.

In our paper, we have largely focused on inter-pulse interferences. As we have showed, they lead to a nearly perfect coherent enhancement of the momentum probability distributions of created particles.
Namely, at certain momenta, the major inter-pulse peaks scale approximately like $N_{\rm rep}^2$ [Eq.~\eqref{pro11}] as compared to the intra-pulse modulations. 
This has been related to adiabatic transitions between different eigenstates of the time-evolution operator characterized by phases $\vartheta_1$ and $\vartheta_2$.
While the gap between these phases has to be sufficiently small for the adiabatic transition to occur,
with increasing the gap the nearly perfect coherence is lost. Other detailed features of the momentum probability distributions 
have also been described using this interpretation.

In closing, we recall that similar -- but fully coherent -- peak structures have been discussed in other strong-field quantum electrodynamics processes such as the Breit-Wheeler 
pair creation~\cite{BWcomb}, the Compton scattering~\cite{Comptoncomb1,Comptoncomb2,Twardy,Comptoncomb3,Comptoncomb4}, and ionization~\cite{Ionizationcomb}. Here, instead, we conclude that a perfect
coherent enhancement of probability distributions of the Sauter-Schwinger pairs can never be reached. This is in contrast to the previous studies of the Sauter-Schwinger 
process~\cite{Akkermans,Li1,Li2}. Note, however, that there are other (i.e., noncoherent) means of amplifying the signal of created pairs. As we have demonstrated here, 
this can be achieved when applying a broader bandwidth pulse. In such case,
more energetic photons participate in pair production; thus, resulting in several orders of magnitude enhancement of the probability distributions as compared to those predicted
in~\cite{Akkermans}. More aspects of such investigations are going to be presented in due course.

\section*{Acknowledgements}

We thank Piotr Chankowski for his careful reading of the manuscript and for providing us with valuable comments. This work is supported by the National Science Centre (Poland) under Grant No. 2014/15/B/ST2/02203.

\appendix

\section{Quantum Vlasov equation}
\label{Vlasov}

The system of equations~\eqref{a40} is equivalent to the quantum Vlasov equation~\cite{Schmidt}, which has been solved in the same context by various authors (see, 
e.g.~\cite{Akal,Li2}).
To see this better, let us introduce the following combinations of the coefficients $c_{\bm p}^{(1)}(t)$ and $c_{\bm p}^{(2)}(t)$ defined by Eqs.~\eqref{a38} and~\eqref{a39},
\begin{align}
f({\bm p},t)&=|c_{\bm p}^{(2)}(t)|^2,\nonumber\\
u({\bm p},t)&=c_{\bm p}^{(1)}(t)[c_{\bm p}^{(2)}(t)]^*+[c_{\bm p}^{(1)}(t)]^*c_{\bm p}^{(2)}(t),\label{vla1}\\
v({\bm p},t)&=\ii \bigl(c_{\bm p}^{(1)}(t)[c_{\bm p}^{(2)}(t)]^*-[c_{\bm p}^{(1)}(t)]^*c_{\bm p}^{(2)}(t)\bigr).\nonumber
\end{align}
Note that each of these functions is real. Moreover, based on Eqs.~\eqref{a28} and~\eqref{a39}, we conclude that $f({\bm p},t)$ defines the temporal probability 
of pair creation in a given eigenmode of the fermionic field, 
${\cal P}(t)$,
\begin{equation}
{\cal P}(t)\equiv f({\bm p},t)=|c_{\bm p}^{(2)}(t)|^2.\label{special}
\end{equation}
Calculating the time derivative of the quantities~\eqref{vla1} and using~\eqref{a40}, we obtain that
\begin{align}
\dot{f}({\bm p},t)&=-\Omega_{\bm p}(t)u({\bm p},t),\nonumber\\
\dot{u}({\bm p},t)&=-2\omega_{\bm p}(t)v({\bm p},t)-2\Omega_{\bm p}(t)[1-2f({\bm p},t)],\label{vla2}\\
\dot{v}({\bm p},t)&=2\omega_{\bm p}(t)u({\bm p},t)\nonumber,
\end{align}
where we have used Eq.~\eqref{a40} and the fact that $|c_{\bm p}^{(1)}(t)|^2=1-f({\bm p},t)$. Since at the initial time $t'$ we had $c_{\bm p}^{(2)}(t')=0$,
the aforementioned system of equations has to be solved with the initial conditions: $f({\bm p},t')=0$, $u({\bm p},t)=0$, and $v({\bm p},t)=0$.

In doing so, we introduce a complex function $\zeta({\bm p},t)=u({\bm p},t)+\ii v({\bm p},t)$ and the following abbreviation, $R({\bm p},t)=2\Omega_{\bm p}(t)[1-2f({\bm p},t)]$.
Then, the last two equations of~\eqref{vla2} can be written in the form of the first-order, inhomogeneous, linear differential equation
for the unknown function $\zeta({\bm p},t)$, 
\begin{equation}
\dot{\zeta}({\bm p},t)=2\ii\omega_{\bm p}(t)\zeta({\bm p},t)-R({\bm p},t).\label{vla3}
\end{equation}
Solving first the homogeneous equation,
\begin{equation}
\dot{\zeta}({\bm p},t)=2\ii\omega_{\bm p}(t)\zeta({\bm p},t),\label{vla4}
\end{equation}
we find out that
\begin{equation}
\zeta({\bm p},t)=C\exp\Bigl(2\ii\int_{t'}^t\dd\tau\omega_{\bm p}(\tau)\Bigr),\label{vla5}
\end{equation}
where $C$ is the integration constant. Varying this constant, $C=C(t)$, and plugging
\begin{equation}
\zeta({\bm p},t)=C(t)\exp\Bigl(2\ii\int_{t'}^t\dd\tau\omega_{\bm p}(\tau)\Bigr),\label{vla6}
\end{equation}
into Eq.~\eqref{vla3}, we arrive at 
\begin{equation}
\dot{C}(t)=-R(t)\exp\Bigl(-2\ii\int_{t'}^t\dd\tau\omega_{\bm p}(\tau)\Bigr).\label{vla7}
\end{equation}
The solution of this equation with the initial condition that $C(t')=0$ is
\begin{equation}
C(t)=-\int_{t'}^t\dd\tau R(\tau)\exp\Bigl(-2\ii\int_{t'}^\tau\dd\sigma\omega_{\bm p}(\sigma)\Bigr).\label{vla8}
\end{equation}
Hence, combining~\eqref{vla6} with~\eqref{vla8}, we obtain that
\begin{equation}
\zeta({\bm p},t)=-\int_{t'}^t\dd\tau R(\tau)\exp\Bigl(2\ii\int_{\tau}^t\dd\sigma\omega_{\bm p}(\sigma)\Bigr),\label{vla9}
\end{equation}
or, equivalently,
\begin{align}
u({\bm p},t)&=-\int_{t'}^t\dd\tau R(\tau)\cos\Bigl(2\int_{\tau}^t\dd\sigma\omega_{\bm p}(\sigma)\Bigr),\label{vla10}\\
v({\bm p},t)&=-\int_{t'}^t\dd\tau R(\tau)\sin\Bigl(2\int_{\tau}^t\dd\sigma\omega_{\bm p}(\sigma)\Bigr).\label{vla11}
\end{align}
Finally, the quantum Vlasov equation is obtained by substituting~\eqref{vla10} into the first equation of~\eqref{vla2},
\begin{align}
\dot{f}({\bm p},t)=2\Omega_{\bm p}(t)\int_{t'}^t&\dd\tau\Omega_{\bm p}(\tau)\bigl(1-2f({\bm p},\tau)\bigr)\nonumber\\
&\times\cos\Bigl(2\int_{\tau}^t\dd\sigma\omega_{\bm p}(\sigma)\Bigr),\label{vla12}
\end{align}
with $\omega_{\bm p}(t)$ and $\Omega_{\bm p}(t)$ defined in Sec.~\ref{theory}.

\section{Two-dimensional unitary matrix}
\label{appendix1}

In this appendix, we introduce the parametrization of a $2\times 2$ unitary matrix, $\op{U}$. In general, such a matrix has the form
\begin{equation}
\op{U}=\begin{pmatrix} U_{11} & U_{12} \cr U_{21} & U_{22} \end{pmatrix},
\label{appa1}
\end{equation}
where, from the unitary condition $\op{U}^{\dagger}\op{U}=\op{I}$, we find that
\begin{align}
|U_{11}|^2+|U_{21}|^2 & =1, \nonumber \\
|U_{12}|^2+|U_{22}|^2 & =1, \nonumber \\
U_{11}^*U_{12}+U_{21}^*U_{22} &=0.
\label{appa2}
\end{align}
This means that a unitary $2\times 2$ matrix can be uniquely defined by four real parameters. 

In addition, we know that any $2\times 2$ unitary matrix has two orthonormal eigenvectors, which we shall denote as $\ket{j}$, and the corresponding complex eigenvalues 
$\lambda_j$ such that $\op{U}\ket{j}=\lambda_j\ket{j}$ ($j=1,2$). The eigenvalues have modulus one, $|\lambda_j|=1$, meaning that they can be represented as 
$\lambda_j=\ee^{-\ii\vartheta_j}$, with arbitrary real angles $\vartheta_j$ defined modulo $2\pi$. For the eigenvectors we can choose
\begin{align}
\ket{1}&=\ee^{\ii\psi_1}\begin{pmatrix} \ee^{-\ii\beta/2}\cos(\gamma/2) \cr
\ee^{\ii\beta/2}\sin(\gamma/2) \end{pmatrix}, \nonumber \\
\ket{2}&=\ee^{\ii\psi_2}\begin{pmatrix} -\ee^{-\ii\beta/2}\sin(\gamma/2) \cr
\ee^{\ii\beta/2}\cos(\gamma/2) \end{pmatrix} ,
\label{appa3ap}
\end{align}
where $0\leqslant\gamma\leqslant\pi$, $0\leqslant\beta<2\pi$, and the global phases $\psi_j$ can be chosen arbitrary and are irrelevant in our further analysis. 
We can also construct two projection operators,
\begin{equation}
\hat{P}_{j}=\ket{j}\bra{j} ,
\label{appa4}
\end{equation}
such that $\op{P}_j^{\dagger}=\op{P}_j$, $\op{P}_j\op{P}_{\ell}=\op{P}_j\delta_{j\ell}$, and $\op{P}_1+\op{P}_2=\op{I}$. As it follows from~\eqref{appa3ap}
and~\eqref{appa4}, they are independent of the global phases $\psi_j$ as
\begin{align}
\hat{P}_1&= \frac{1}{2}\begin{pmatrix}
1+\cos\gamma & \ee^{-\ii\beta}\sin\gamma \cr \ee^{\ii\beta}\sin\gamma & 1-\cos\gamma
\end{pmatrix} ,
\nonumber \\
\hat{P}_2&= \frac{1}{2}\begin{pmatrix}
1-\cos\gamma & -\ee^{-\ii\beta}\sin\gamma \cr -\ee^{\ii\beta}\sin\gamma & 1+\cos\gamma
\end{pmatrix} .
\label{appa5}
\end{align}
Using the spectral decomposition for the operator $\op{U}$,
\begin{equation}
\hat{U}=\lambda_1\op{P}_1+\lambda_2\op{P}_2 ,
\label{appa6}
\end{equation}
we can write it down explicitly,
\begin{equation}
\hat{U}= \ee^{-\ii\vartheta_0}
\begin{pmatrix}
\cos\vartheta+\ii\sin\vartheta\cos\gamma & \ii\ee^{-\ii\beta}\sin\vartheta\sin\gamma
\cr
-\ii\ee^{\ii\beta}\sin\vartheta\sin\gamma & \cos\vartheta-\ii\sin\vartheta\cos\gamma
\end{pmatrix} ,
\label{appa7}
\end{equation}
with $\vartheta_0=(\vartheta_2+\vartheta_1)/2$ and $\vartheta=(\vartheta_2-\vartheta_1)/2$. Hence, the four real angles $0\leqslant \vartheta_0,\vartheta < 2\pi$, $\beta$, 
and $\gamma$ parametrize an arbitrary $2\times 2$ unitary matrix.

One can relate these angles to the elements of the matrix $\op{U}$ in the form given by Eq.~\eqref{appa1}. For this purpose, we determine the eigenvalues of the matrix~\eqref{appa1},
\begin{equation}
\lambda_1=\frac{U_{11}+U_{22}+\sqrt{\Delta}}{2}, \quad \lambda_2=\frac{U_{11}+U_{22}-\sqrt{\Delta}}{2},
\label{appb2}
\end{equation}
where
\begin{equation}
\Delta=(U_{11}-U_{22})^2+4U_{12}U_{21},
\label{appb1}
\end{equation}
and where we choose $\sqrt{\Delta}$ such that $\mathrm{Re}\sqrt{\Delta} >0$. Hence, the eigenvector corresponding to $\lambda_1$ is found,
\begin{equation}
\ket{1}=\frac{\ee^{\ii\chi_1}}{\sqrt{|\lambda_1-U_{11}|^2+|U_{12}|^2}}
\begin{pmatrix} U_{12} \cr \lambda_1-U_{11} \end{pmatrix},
\label{appb3}
\end{equation}
where $\chi_1$ is an unknown phase. Its comparison with~\eqref{appa3ap} leads to a set of two equations,
\begin{align}
\psi_1-\beta/2=&\arg(\ee^{\ii\chi_1}U_{12})=\chi_1+\arg(U_{12}), \\
\psi_1+\beta/2=&\arg(\ee^{\ii\chi_1}(\lambda_1-U_{11}))=\chi_1+\arg(\lambda_1-U_{11}), \nonumber
\label{align}
\end{align}
where $\arg(z)$ is the phase of the complex number $z$. Next, subtracting both sides of these equations, we get
\begin{equation}
\beta=\arg(\lambda_1-U_{11})-\arg(U_{12}) \mod 2\pi .
\label{appb5}
\end{equation}
Furthermore, from the projection matrix $\op{P}_1$, we obtain
\begin{equation}
\frac{1+\cos\gamma}{2}=\frac{|U_{12}|^2}{|U_{12}|^2+|\lambda_1-U_{11}|^2},
\label{appb6}
\end{equation}
and, hence,
\begin{equation}
\gamma=\arccos\Bigl(\frac{|U_{12}|^2-|\lambda_1-U_{11}|^2}{|U_{12}|^2+|\lambda_1-U_{11}|^2}\Bigr).
\label{appb7}
\end{equation}
Finally,
\begin{equation}
\vartheta_j=-\arg(\lambda_j),\quad j=1,2.
\label{appb8}
\end{equation}
In this way we have uniquely determined all relevant angles.

\end{document}